\documentclass[twocolumn,showpacs,preprintnumbers,amsmath,amssymb,pra]{revtex4}
\usepackage{graphicx}
\newcommand{\bm}[1]{ \mbox{\boldmath $#1$}  }

\begin{document}

\title{Effective Potential for Ultracold Atoms at the Zero-Crossing of a Feshbach Resonance}

\author{N.T. Zinner}
\affiliation{ Department of Physics, Harvard University, Cambridge,
  Massachusetts 02138, USA}
\altaffiliation{Present address: 
Department of Physics
and Astronomy, University of Aarhus,
DK-8000 {\AA}rhus C, Denmark}

\date{\today}

\hyphenation{Fesh-bach}

\begin{abstract}
We consider finite-range effects when the scattering length goes to zero near a magnetically controlled 
Feshbach resonance. The traditional effective-range expansion is badly behaved at this point
and we therefore introduce an effective potential that reproduces the full $T$-matrix. To lowest order
the effective potential goes as momentum squared times a factor that is well-defined as the scattering
length goes to zero. The potential turns out to be proportional to the background scattering 
length squared times the background effective range for the resonance. 
We proceed to estimate the applicability and relative importance of this potential for Bose-Einstein condensates 
and for two-component Fermi gases 
where the attractive nature of the effective potential can lead to collapse above a critical particle number
or induce instability toward pairing and superfluidity. For broad Feshbach resonances the higher-order effect is
completely negligible. However, for narrow resonances in tightly confined samples signatures might be 
experimentally accessible. This could be relevant for sub-optical wavelength microstructured traps
at the interface of cold atoms and solid-state surfaces.
\end{abstract}
\maketitle

\section{Introduction}
Cold atomic gases have enjoyed many great successes since the first 
realizations of Bose-Einstein condensates in the mid-nineties \cite{bloch2008} 
Ensembles of ultracold atomic gases can be manipulated in magnetic or optical
trap geometries and in lattice setups, effectively mimicking the structure of 
real materials and teaching us about their properties. In particular, extreme
control can be exercised over the atom-atom interactions through the 
use of Feshbach resonance \cite{chin2010}. Tuning the system into the 
regime of resonant two-body interactions provides a controlled way of 
studying strongly correlated dynamics which is believed to be crucial 
for material properties such as high-temperature superconductivity or
giant magnetoresistance.

Recently there has been extended interesting in weakly interacting Bose-Einstein condensates for use as an atomic
interferometer \cite{fattori2008a} and also to probe magnetic dipolar interactions in condensates \cite{fattori2008b}.
This work was based on $^{39}$K atoms where a broad Feshbach resonance exists at a magnetic field strength of $B_0=402.4$G
\cite{errico07} which allows a large tunability of the atomic interaction in experiments \cite{roati2007}. Similar
tunability has also been reported in a condensate of $^{7}$Li \cite{pollack2009}. 
The atomic interaction can be reduced by tuning the scattering length, $a$, to zero, also known as zero-crossing. In a 
Gross-Pitaevskii mean-field picture we can thus neglect the usual non-linear term proportional to $a$. 
The question is then what other interactions are relevant. As shown in \cite{fattori2008b}, the magnetic dipole will contribute here. 

In the Gross-Pitaevskii picture we might also ask whether higher-order terms in the interaction can contribute around
zero-crossing. Recently it was shown that effective-range corrections can in fact influence the stability
of condensates around zero-crossing \cite{gao2003,zinner2009,thoger2009}. The Feshbach resonances used thus far in experiments have typically 
been very broad, and as a result the effective range, $r_e$, will be small, rendering the higher-order terms negligible. 
However, around narrow resonances this is not necessarily the case and finite-range corrections are not necessarily negligible. 

For the two-component Fermi gas, there has been increased interest in producing a cold atom
analog of the celebrated Stoner model of ferromagnetism \cite{stoner1933} which applies to 
repulsively interacting fermions. Theoretical proposals indicate that this should be 
possible \cite{ferrotheory} and an MIT experiment subsequently announced indications of 
the ferromagnetic transition \cite{jo2009}. The results caused controversy since the 
spin domains were not resolved \cite{hui2009}. A later experiment in the same 
group did not find evidence of the ferromagnetic transition \cite{sanner2011}.
However, these studies consider broad Feshbach resonances and the situation with
narrow resonances is less clear. One can imagine that finite-range corrections 
could play a role in driving the phase transition. In fact, a recent experiment in
Innsbruck \cite{innsbruck2011} has found increased lifetimes of the repulsive
gas in the strongly imbalanced case, providing hope that decay into molecules
can be controlled and ferromagnetism can be studied. 

The systematic inclusion of finite-range effects through derivative terms in zero-range models was begun in 
the study of nuclear matter decades ago \cite{skyrme1956}. Later on the intricacies of the cut-off problems
that arise in this respect was considered by many authors both for the relativistic and non-relativistic case 
(see \cite{phil98} for discussion and references). In the context of cold atoms and Feshbach resonances, 
we need to use a two-channel model \cite{bruun05} in order to take the lowest order finite-range term into account.
Similar models were already introduced in \cite{kokkelmanns02} and denoted resonance models 
(see f.x. Ref.~\cite{braaten08} for a comprehensive review of scattering models for ultracold atoms). We note that 
whereas resonance models treat the closed-channel molecular state as a point boson the model of Ref.~\cite{bruun05} 
treats the molecule more naturally as a composite object of two atoms. In the end the parameters
of the two models turn out to be similarly related to the physical parameters of Feshbach resonances 
(see for instance the discussion of resonance models in Ref.~\cite{braaten08}). 

\begin{figure}[ht!]
  \includegraphics*[angle=0,scale=0.55]{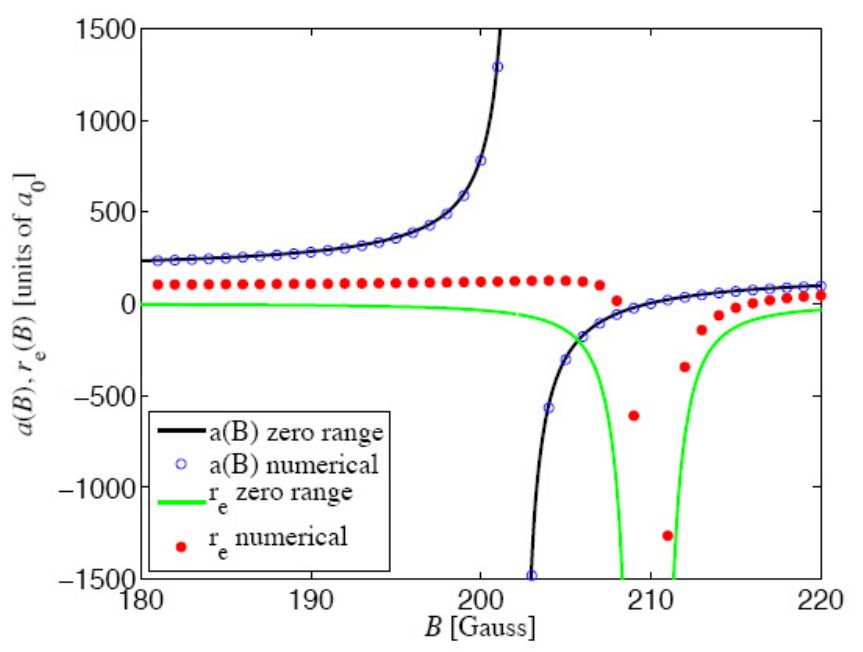}
  \caption{Scattering length and effective range for the $s$-wave scattering of fermionic $^{40}$K atoms around the Feshbach resonance at $B_0=202.1$ G demonstrating the divergence in a coupled-channel calculation (symbols) \cite{nygaard06} and in a zero-range model (full lines). The difference in the zero-range and coupled-channel models is caused by the presence of a bound state close to threshold in the open channel.}
  \label{fig-scat}
\end{figure}

In Fig.~(\ref{fig-scat}) we show calculations of scattering length and effective range for the Feshbach resonance 
at $B=202.1$ G in $^{40}$K in both a coupled-channel model \cite{nygaard06} and in the zero-range model discussed here. 
We see the effective range being roughly constant at resonance and then start to diverge at zero-crossing. The 
zero-range model provides a good approximation to the full calculations and for many-body purposes it is 
preferable due to its simplicity.

Whereas the earlier work of Ref.~\cite{kokkelmanns02} considered the regime close to the resonance, we will 
be exclusively concerned with zero-crossing. To our knowledge the intricacies of this region have not 
been addressed in the literature in the context of Feshbach resonances.
Around zero-crossing the Feshbach model turns out to have a badly behaved effective-range expansion. 
The parameters obtained from the effective-range expansion should therefore be
used with extreme caution as the series is divergent at this point.
However, as we show in this article, the 
finite-range corrections obtained from the full $T$-matrix at low momenta via an effective potential turn 
out to be the same as one would naively expect based on the effective-range expansion.
After introducing the effective potential
we consider its applicability and importance in the case of Bose-Einstein condensates 
and for two-component Fermi gases where the attractive nature of the effective interaction
at zero-crossing could lead to collapse above a certain critical particle number or to pairing 
instability and superfluidity. In general, we find that tight external confinement is a necessary condition for 
the higher-order effects to dominate the magnetic dipole interaction and be experimentally observable.

\section{Two-Channel Model}
We consider a two-channel $s$-wave Feshbach model with zero-range interactions \cite{bruun05} for which the on-shell open-open channel $T$-matrix as
a function of magnetic field, $B$, is
\begin{equation}\label{full}
  T_{oo}(B)=\frac{\frac{4\pi \hbar^2}{m} a_{bg}}{\left( 1+\frac{\Delta\mu \Delta B}{\frac{\hbar^2 q^2}{m}-\Delta\mu(B-B_0)  }\right)^{-1}+i a_{bg}q},
\end{equation}
where $\Delta\mu$ is the difference between the magnetic moments in
the open and closed channel, $q$ is the relative momentum of the
atoms of mass $m$, $a_{bg}$ is the scattering length away from the resonance at 
magnetic field $B_0$, and $\Delta B$ is the width of the resonance.
We can compare this to the standard vacuum expression for the
$T$-matrix in terms of the phase-shift given by
\begin{equation}\label{tvac}
  T_v=\frac{\frac{4\pi\hbar^2 }{m}a }{-qa\cot\delta(q)+iaq}.
\end{equation}

Typically, one has the low energy expression $-q\cot\delta(q)\to -1/a$
which implies that 
\begin{equation}
  T_v\to\frac{\frac{4\pi\hbar^2 }{m}a }{1+iaq}\to 0.
\end{equation}
However, as we now discuss, for the realistic two-channel $T$-matrix for
Feshbach resonances, the quantity $-q\cot\delta(q)$ is not well-defined, and
the conclusion that the $T$-matrix vanishes at zero-crossing is only true for 
zero momentum, $q=0$, as we now discuss.

From Eqs.~\eqref{full} and \eqref{tvac} we obtain the relation for the phase-shift
\begin{equation}\label{fulleff}
q\cot\delta(q)=\frac{-1}{a_{bg}}\left( 1+\frac{\Delta\mu \Delta B}{\frac{\hbar^2 q^2}{m}-\Delta\mu(B-B_0)  }\right)^{-1}.
\end{equation}
We now expand the right-hand side in powers of $q$ as is usually done in an effective-range expansion. This yields
\begin{align}\label{expand}
q\cot\delta(q)=&\frac{-1}{a(B)}&\nonumber\\&+\sum_{n=1}^{\infty}\frac{-1}{a_{bg}}\left[\frac{-a_{bg}r_{e0}}{2}\right]^{n}\left[\frac{a_{bg}}{a(B)}-1\right]^{n+1}q^{2n},&
\end{align}
where $a(B)=a_{bg}\left(1-\frac{\Delta B}{B-B_0}\right)$ is the common parametrization from 
single-channel models and
$r_{e0}=-2\hbar^2/(m\Delta B\Delta\mu a_{bg})$ is the background value of the effective range around the resonance. From Eq.~\eqref{expand} we can now
read off all coefficients in an effective-range expansion with their full $B$-field dependence. For instance, the effective range 
is given simply by $r_e=r_{e0}\left[\tfrac{a_{bg}}{a}-1\right]^2$, which is divergent when $a(B)\rightarrow0$. We also clearly see that all the
other coefficients are divergent in that limit. This is signaled also before doing the full expansion in $q$ as the
first term in Eq.~\eqref{expand} diverges at zero-crossing. However, in effective potentials derived from the $T$-matrix
these problems are not transparent as the lowest order coefficient is proportional to $a(B)$ (see Eq.~\eqref{seff}).
Below we will discuss what kind of constraints this introduces on the 
applicability of the effective-range expansion near zero-crossing. We note that similar issues were briefly discussed in a 
different context in \cite{massignan06} where an equivalent to Eq.~\eqref{teq} below was obtained.

Let us first consider the low-$q$ limit and compare the full $T$-matrix with the effective-range expansion as zero-crossing
is approached. Taking the low-$q$ limit of Eq.~\eqref{fulleff} at zero-crossing where $\Delta B/(B-B_0)=1$, we find
\begin{align}
q\cot\delta(q)\rightarrow \frac{-1}{a_{bg}}-\frac{\Delta\mu\Delta B}{\frac{\hbar^2q^2}{m}},
\end{align}
which diverges as $q^{-2}$. 
Therefore the coefficients of the expansion in Eq.~\eqref{expand} must necessarily diverge in order to retain any hope of describing the low-$q$ behavior. Furthermore, since the
expansion is an alternating series and therefore slowly converged, we also conclude that many terms must be retained for a fair approximation at very small but non-zero $q$. The same conclusion can be reached by considering the radius of convergence of Eq.~\eqref{expand}, which we find by locating the pole in Eq.~\eqref{fulleff} at $\hbar^2 q^2/m=\Delta\mu(B-B_0-\Delta B)$. This radius indeed goes to zero at zero-crossing. We are thus forced to conclude that the effective-range expansion breaks down near zero-crossing.

\subsection{Effective Potential at Zero-crossing}
Since the effective-range expansion is insufficient we consider the full $T$-matrix in the low-$q$ limit at zero-crossing. To lowest order
we have
\begin{equation}\label{teq}
T_{oo}(B=B_0+\Delta B)=-\frac{4\pi\hbar^2a_{bg}}{m}\frac{\hbar^2 q^2}{m\Delta\mu\Delta B}+O(q^4).
\end{equation}
Using the expression for $r_{e0}$, this can be written 
\begin{equation}
\frac{4\pi\hbar^2}{m}\frac{a_{bg}^{2}r_{e0}}{2}q^2.
\end{equation}
Knowing the $T$-matrix at low $q$ we can now proceed to find an effective low-$q$ potential through the Lippmann-Schwinger equation
\begin{align}
V=T-TG_0 V,
\end{align}
where $G_0=(E-H_0+i\delta)^{-1}$ is the free space Green's function \cite{pet02}. This equation can be solved for $T(q,q')\propto{q}^2+{q'}^2$ (the symmetrized version of the full $T$-matrix) in an explicit cut-off approach \cite{phil98,pet02} and then be expanded to 
order $q^2$ for consistence with the input $T$-matrix. In the long-wavelength limit we can take the cut-off to 
zero \cite{pet02} and for the on-shell effective potential we then obtain the obvious answer
\begin{align}\label{qpot}
V(q)=\frac{4\pi\hbar^2}{m}\frac{a_{bg}^{2}r_{e0}}{2}q^2
\end{align}
in momentum space. The effective potential in real-space is now easily found by canonical substitution ($\bm q\rightarrow -i\nabla$) and appropriate symmetrization \cite{roth01}. We have
\begin{align}\label{effpot}
V(\bm r)=-\frac{4\pi\hbar^2}{m}\frac{a_{bg}^{2}r_{e0}}{2}\frac{1}{2}\left[\overleftarrow{\nabla}_{\bm r}^2\delta(\bm r)
  +\delta(\bm r)\overrightarrow{\nabla}_{\bm r}^2\right].
\end{align}
Notice that the Lippmann-Schwinger approach is non-perturbative as opposed to the perturbative energy shift method \cite{roth01,pet07}.

\subsection{Comparison to Effective-Range Expansion and Energy-Shift Method}
Away from zero-crossing one can easily relate the effective-range expansion to an effective potential through the perturbative energy shift method \cite{phil98,roth01,pet07}. To second order the $s$-wave effective potential is
\begin{align}\label{seff}
V(\bm r)=\frac{4\pi\hbar^2 a}{m}\left[\delta(\bm r)+\frac{g_2}{2}\left(\overleftarrow{\nabla}_{\bm r}^2\delta(\bm r)
  +\delta(\bm r)\overrightarrow{\nabla}_{\bm r}^2\right)\right],
\end{align}
where the first term is the effective interaction usually employed in mean-field theories of cold atoms \cite{pet02}. In terms of $a$ and $r_e$, we have $g_2=a^2/3-ar_e/2$ \cite{roth01,pet07} with the field-dependent $a=a(B)$ and $r_e=r_e(B)$. 

At zero-crossing the first term in Eq.~\eqref{seff} vanishes and one might expect the second term to vanish as well. However, in the naive effective-range expansion of the two-channel model discussed above we saw that $r_e$ diverges as $a^{-2}$ and we therefore have
\begin{align}\label{const}
\lim_{a\rightarrow 0} ag_2=-\frac{a_{bg}^2 r_{e0}}{2}.
\end{align}
In particular, if we for a moment ignore $q^4$ terms in the effective-range expansion, we recover exactly the same effective potential as in Eq.~\eqref{effpot} at zero-crossing. The finite limiting result in Eq.~\eqref{const} shows that the potential in Eq.~\eqref{seff} is 
well-defined as $a\rightarrow 0$, provided that appropriate regularization and renormalization is performed. Eq.~\eqref{seff} thus applies equally well at resonance ($a\rightarrow\infty$) where the gradient terms are small and at zero-crossing where the lowest order delta function term is unimportant. It is thus a well-defined effective potential over the entire range of a Feshbach resonance.

We therefore see that even though the effective-range expansion has divergent coefficients at zero-crossing, the lowest order does in fact give the same effective potential as the full $T$-matrix if we apply it naively.  The effective-range expansion should thus be viewed as an asymptotic series. However, we cannot use the effective-range expansion to estimate the validity of the second order effective potential since the radius of convergence goes to zero at zero-crossing as discussed above.

The two-channel model in Eq.~\eqref{full} compares well with a coupled-channel calculation \cite{nygaard06}
as shown in Fig.~\ref{fig-scat}. It also compares well to other scattering models \cite{marcelis04,marcelis06}
that include finite-range effects. In fact, the model used here compares well with the analytical models
of Ref.~\cite{marcelis04} when $a(B)$ and $r_e(B)$ have the field-dependence introduced above. This can 
be seen for instance in Fig.~12 of Ref.~\cite{marcelis04}, although a difference is that our $a(B)$ and
$r_e(B)$ are parametrization and not taken from coupled-channels values as in Ref.~\cite{marcelis04} 
(our Fig.~\ref{fig-scat} quantifies the difference which is largest on $r_e(B)$). However, here we are 
concerned with the behavior when $a(B)\to 0$ in the context of Feshbach resonances which is not 
addressed in Refs.~\cite{marcelis04,marcelis06}. We note that the resonance models of Ref.~\cite{kokkelmanns02}
and the two-channel and resonance models in Ref.~\cite{braaten08} are very similar to the model employed
here, but again those references do not consider the specific problems arising when $a(B)\to 0$. In addition and 
in contrast to previous discussions,
here we construct appropiate zero-range (pseudo)-potentials that work around zero-crossing.

\section{Relation to Experiments}
Above we only retained terms of order $q^2$ in the full $T$-matrix. We now estimate the energy regime in which this expression is valid. Demanding 
that the $q^4$ term be smaller than the $q^2$ term gives the criterion
\begin{align}\label{cond}
\frac{\hbar^2q^2}{m} \ll \frac{\hbar^2}{m|a_{bg}r_{e0}|}.
\end{align}
We relate this condition to recent experiments with bosonic condensates of $^{39}$K working around zero-crossing \cite{fattori2008a}. The 
resonance used there is very broad ($\Delta B=-52$G) with $a_{bg}=-29a_0$ and $r_{e0}=-58a_0$ ($a_0$ is the Bohr radius). The right-hand side of Eq.~\eqref{cond} 
is  $2.3\cdot 10^{-7}$ eV, corresponding to a temperature of about 3 mK. Since the experiments are performed at much lower temperatures the
approximation above is certainly valid. However, as $a_{bg}$ and particularly $r_{e0}$ is small, the front factor in Eq.~\eqref{effpot} is
also small. The relevant scale of comparison is the outer trap parameter $b$ \cite{zinner2009} which is typically of order $1\mu$m, 
yielding a vanishing ratio $|a_{bg}^{2}r_{e0}|/b^3\sim 10^{-9}$. For broad Feshbach resonances the higher-order interactions can thus be safely ignored. For very narrow resonances the situation potentially changes as 
$r_{e0}$ can be very large and
make the potential in Eq.~\eqref{effpot} important. As an example, we consider the narrow resonance in $^{39}$K at $B_0=25.85$G with $\Delta B=0.47$G, $a_{bg}=-33a_0$, and $r_{e0}=-5687a_0$ \cite{errico07}. The right-hand side of Eq.~\eqref{cond} is now $2\cdot 10^{-9}$ eV, corresponding to 24 $\mu$K. This is again much higher than experimental temperatures. A more careful argument can be made from the energy per particle of the non-condensed cloud. Ignoring the trap, we have $E/N=0.770k_B T_c (T/T_c)^{5/2}$ ($T_c$ is the critical temperature) \cite{pet02}. For a sample of $3\cdot 10^4$ a critical temperature of 100 nK was reported in \cite{roati2007}. Using this $T_c$ we find that $T\ll 900$nK for Eq.~\eqref{cond} to hold. Again this is within the experimental regime. The effective potential approach should therefore be applicable around zero-crossing for narrow resonances. However, even with this narrow resonance we find $|a_{bg}^{2}r_{e0}|/b^3\sim 10^{-7}$ and the 
effect is still completely negligible.

In order to increase the relevance of the higher-order term, we now consider some very narrow resonances that have been found in $^{87}$Rb. In particular, the resonance at $B_0=9.13$G \cite{widera04} which was recently utilized in nonlinear atom interferometry \cite{gross10}. We have 
$\Delta B=0.015$G, $a_{bg}=99.8a_0$, and $\Delta\mu=2.00\mu_B$ \cite{chin10}, which gives $r_{e0}=-19.8\cdot 10^3a_0$ and a ratio $|a_{bg}^{2}r_{e0}|/b^3=2.92\cdot 10^{-5}(1\mu\text{m}/b)^3$. A trap length of $b\sim 0.5\mu$m as used in \cite{gross10} would thus yield $10^{-4}$ and demonstrates that higher-order corrections can safely be neglected. For a ratio of 1 we need $b\sim 0.03\mu$m which is unrealistically small in current traps or optical lattices. However, a resonance of width $\Delta B=0.0004$G is known in the same system at $B_0=406.2$G \cite{marte02} with $a_{bg}=100a_0$ and $\Delta\mu=2.01\mu_B$ \cite{chin10}. In this case we find $r_{e0}=-7.4\cdot 10^5a_0$ and a much more favorable ratio of $|a_{bg}^{2}r_{e0}|/b^3=0.001(1\mu\text{m}/b)^3$. Here we see that a ratio of 1 is achieved already for $b\sim 0.1\mu$m which not far off from tight traps or optical lattice dimensions. In terms of temperature we still have to be in the ultralow regime of $T\lesssim 30$nK according to Eq.~\eqref{cond} for the latter resonance.

Consider now a fermionic two-component system where $s$-wave interactions are dominant. Since we have $r_{e0}<0$ for
all Feshbach resonances \cite{chin10}, the effective potential in Eq.~\eqref{qpot} is attractive and the system could potentially
be unstable toward a paired state or become unstable to collapse above a critical particle number. For simplicity we will use the semi-classical
Thomas-Fermi approach to describe a gas with equal population of the two components and estimate the critical particle number. Assuming an isotropic trapping potential with length scale $b=\sqrt{\hbar/m\omega}$ where $\omega$ is the trap frequency, the ground-state density, $\rho(\bm x)$, can be found by minimization and satisfies
\begin{align}\label{eom}
\left[\frac{\mu}{\hbar\omega} -\frac{1}{2}\left(\frac{\bm x}{b}\right)^2\right]=&\frac{1}{2}(k_F(\bm x)b)^2
-\frac{4}{30\pi}\alpha(k_F(\bm x)b)^5,
\end{align}
where $\rho(\bm x)=k_{F}(\bm x)/6\pi^2$ and $\alpha=a_{bg}^{2}|r_{e0}|/b^3$. The maximum allowed momentum and chemical potential, $\mu$, is
found by solving for the turning point of the right-hand side of Eq.~\eqref{eom} which gives
\begin{align}
k_{max}b=\left[\frac{3\pi}{2\alpha}\right]^{1/3}\quad\text{and}\quad\mu_{max}=\frac{3}{10}\hbar\omega(k_{max}b)^2.
\end{align}
We can now compare this $k_{max}$ to the value obtained from the non-interacting density within the Thomas-Fermi approximation at the center
of the trap. In terms of the number of particles in each component, $N$, at the center of the trap we have $k_{F}(0)b\approx 1.906N^{1/6}$ \cite{pet02}. By equating these two expression we obtain an estimate for the critical number of particles, $N_{max}$. Inserting the 
relevant units, we have
\begin{align}
N_{max}=2\cdot 10^{25}\left(\frac{a_0}{a_{bg}}\right)^{4}\left(\frac{a_0}{r_{e0}}\right)^{2}\left(\frac{b}{1\,\mu\text{m}}\right)^{6},
\end{align}
where $a_0$ is the Bohr radius. We note that the scaling $N_{max}\propto \alpha^{-2}$ can also be obtained by considering the point at which the monopole mode becomes unstable.

Typical numbers for common fermionic species $^{6}$Li or $^{40}$K in the lowest hyperfine states \cite{chin10} lead to $N_{max}\sim 10^{12}$ for $b=1\,\mu\text{m}$. This is of course a huge number and experiments are well within this limit. Even if one reduced the trap length by a factor of ten and made the presumably unrealistic assumption that the particle number remains the same we still have
$N\ll N_{max}$. The reason is that the $s$-wave Feshbach resonances utilized in the two-component gases are generally broad in order to study the 
universal regime. If we consider the narrow resonance at $B_0=543.25$G in $^{6}$Li \cite{strecker03} with $\Delta B=0.1$G, $a_{bg}=60a_0$, and $\Delta\mu=2.00\mu_B$ \cite{chin10}, we have $N_{max}\sim 2\cdot 10^{13} (b/1\mu\text{m})^6$. This is somewhat better but we still need $b\sim 0.06\mu$m to get to an experimentally relevant $N_{max}\sim 10^6$. We have to conclude that higher-order $s$-wave interactions are highly unlikely to be observable through monopole instabilities. In light of this it seems better to consider $p$-wave resonances which are much more narrow in general. However, also here extremely small trap sizes appear necessary \cite{zinner2009b}.

The instability toward Cooper pairing around zero-crossing can also be estimated in simple terms. In general the critical temperature is $T_c\sim T_F \exp(-1/N_0 |U|)$, where $N_0=mk_F(0)/2\pi^2\hbar^2$ is the density of states at the Fermi energy in the trap center and $U<0$ is a measure of the attraction. For the latter we use the effective potential in momentum space from Eq.~\eqref{qpot} and make the assumption that $q\sim k_F(0)$. 
Using the expression for $k_F(0)$ in terms of $N$ above, we find 
\begin{align}
\frac{1}{N_0|U|}=\frac{1.5\cdot 10^{12}}{\sqrt{N}}\left(\frac{b}{1\,\mu\text{m}}\right)^3\left(\frac{a_0}{a_{bg}}\right)^2\frac{a_0}{|r_{e0}|}.
\end{align}
For broad resonances in $^{6}$Li or $^{40}$K this exponent is of order $10^3$ and $T_c$ is thus vanishingly small. However, the scaling with trap size can help and if we imagine reducing to $b=0.1\,\mu\text{m}$, we find $T_c\lesssim 0.5T_F$ for $N=10^6$ atoms. For the narrow resonance in $^{6}$Li discussed above, we find that $T_c\sim 0.5T_F$ with $N=10^6$ can be achieved for $b\sim 0.5\mu$m and $T_c\sim 0.1T_F$ for $N=10^5$. Thus there may be a possibility to reach the pairing instability near zero-crossing if high particle numbers can be cooled in tight traps and narrow resonances are used.

While the sub-optical wavelength trapping sizes needed for the above effects to be large are not achievable with 
typical optical or magnetic traps or optical lattice setups, they could potentially be 
reached via hybrid setups where atoms are trapped near a surface. Inspired by surface plasmon 
subwavelength optics \cite{barnes2003}, nanoscale trapping for neutral atoms has been 
studied \cite{murphy2009,chang2009}, and micropotential traps with width less than 100 
nanometer ($<0.1\mu$m) are within reach \cite{stehle2011}. In these very tightly confined
systems, it is very likely that finite-range effects could be enhanced. Devices that 
provide an interface between atoms and solid-state systems are under intense study at 
the moment, and our considerations here imply that finite-range corrections should
be considered when the scattering length is tuned close to zero.

\subsection{Dipole-Dipole Interactions}
The discussion above ignores the dipole-dipole interaction discussed in the introduction which will compete against the higher-order effective potential from the Feshbach resonance. A simple estimate can be made along the lines of the discussion in \cite{pet02}. The external trapping potential is the characteristic scale of spatial variations and we thus find a ratio, $r$, of 
magnetic dipole-dipole, $U_{md}$, to higher-order $s$-wave zero-range interaction strength, $U_{2}$, which can be written as
\begin{align}
r=\frac{U_{md}}{U_{2}}=\frac{a_0 b^2}{a_{bg}^{2}|r_{e0}|}=
35.7\left[\frac{b}{1\mu\text{m}}\right]^2 \left[\frac{100a_0}{a_{bg}}\right]^2
\frac{1000a_0}{|r_{e0}|}.
\end{align}
For $r<1$ the higher-order interaction term will therefore dominate the magnetic dipole term. For the case of narrow resonances in $^{87}$Rb discussed above we find $r\sim 0.11(b/1\mu\text{m})^2$ for the resonance at $B_0=9.13$G and $r\sim 0.05(b/1\mu\text{m})^2$ for the one at $B_0=406.2$G. For the narrow resonance in $^{6}$Li at $B_0=543.25$G we find $r\sim 1.4 (b/1\mu\text{m})^2$. These ratios clearly indicate that 
magnetic dipole-dipole interactions can be suppressed relative to higher-order zero-range terms for narrow Feshbach resonances and standard trap sizes. This domaninace becomes even stronger for the tight traps needed for the realization of the effects discussed above and we thus conclude that interference of the magnetic dipole-dipole term is not a major concern.

\section{Conclusions and Outlook}
In this article we have discussed the effective potential around a Feshbach resonances as the scattering length is tuned to zero and finite-range corrections become important. We showed that the effective-range expansion is badly behaved and the effective potential most be defined from the $T$-matrix.
We have demonstrated that the low momenta effective potential obtained from the full $T$-matrix agrees with one obtained naively from the
effective-range expansion when the scattering length goes to zero. Thus even though the effective-range expansion has divergent coefficients at zero-crossing the first terms of the associated effective potential yield consistent results. We then estimated the effects of the terms
on different condensates. Since the effective potential at zero-crossing is attractive it may induce various instabilities which we considered for the case of a two-component Fermi gas under harmonic confinement.

For the broad Feshbach resonances used in current experiments the effective potential discussed here are negligible and the dipole-dipole interaction dominates completely at zero-crossing. However, for narrow resonances in very tightly confined systems some of the effects might be detectable. 
In particular, future generations of microtraps with sub-optical wavelength trap sizes using surface plasmons could be small enough to make 
finite-range effects important.
The competing dipole interaction is small for narrow resonances in tight confinement. However, it is conceivable that effects of spherically symmetric higher-order terms could be separated from dipolar effects which change with system geometry \cite{fattori2008b}.

Small trapped Fermi systems have recently become an experimental reality with particle numbers ranging from two to ten \cite{serwane2011}. 
For two atomic fermions with different internal states, the system turns out to be well described by the analytic zero-range model of Busch {\it et al.} \cite{busch1998,stoferle2006,blume2002}, and similarly for three fermions \cite{werner2006}. Effective-range corrections to these
results have also been studied \cite{idzia2006,suzuki2009,peng2011,zinner2011}. Mesoscopic Fermi systems (less than about 50 particles) have been studied in harmonic traps using a number of numerical methods \cite{numerics}, with particular emphasis on the unitary regime where the scattering length diverges. It would be interesting to investigate the situation also around zero-crossing of a narrow resonance where the effective range is sizable. A preliminary study along this line for three bosons is discussed in Ref.~\cite{zinner2011boson}.

Another interesting direction of future work is the study of the contact introduced by Tan \cite{tan2008,braaten2008,zhang2009,combescot2009,werner2010,barth2011,valiente2011,valiente2012,manuel2012} to describe the universal behavior of strongly interacting quantum gases at a broad resonance where the range corrections are negligible, for instance through the tail of the momentum distribution which is predicted to behave as $C/k^4$, where $C$ is the contact and $k$ the momentum of a single particle. The relations found by Tan \cite{tan2008} have subsequently been confirmed experimentally in three dimensions \cite{stewart2010,kuhnle2010,kuhnle2011}. 
While the contact originally pertains to two-body correlations, signatures of three-body physics in momentum distributions has also 
been studied both theoretically \cite{werner2010,kang2011,castin2011,helfrich2011,bellotti2011} and experimentally \cite{wild2011}.
While a few studies have considered the universal behavior when including the effective range term \cite{platter2008,werner2008}, 
it would be very interesting to consider the regime around zero-crossing for a narrow resonance where the background 
effective range parameter.

\paragraph{Acknowledgments}
The author would like to thank Martin Th{\o}gersen for very fruitful collaborations. Correspondence with Georg Bruun about two-channel models is highly appreciated. I am grateful to Nicolai Nygaard for discussions and for producing Fig.~(\ref{fig-scat}). I acknowledge the hospitality of the 
Niels Bohr Institute, Blegdamsvej 17, DK-2100 Copenhagen {\O}, Denmark. This work was supported by the Villum Kann Rasmussen foundation.

\end{document}